# The Nature and Evolution of Absorption–Selected Galaxies

Charles C. Steidel[1]

[1] MIT, Physics Department, Cambridge, MA, USA

**Abstract.** We present results of surveys for high redshift galaxies selected by their having produced detectable Mg II and Lyman limit absorption in the spectra of background QSOs. We discuss the properties of the absorbing galaxies, the connection between galaxy properties and absorption line signatures, and how a combination of QSO absorption line and conventional faint galaxy techniques can be used to study field galaxy evolution to very large redshifts.

## 1 Introduction

The subject of QSO absorption lines seems to be reaching a remarkable level of maturity, where the statistics of the various classes of systems are well–known over a very large range of redshifts, and efficient high resolution spectrographs on large telescopes are making feasible (and relatively routine) the accumulation of detailed information for individual absorption systems. Nevertheless, the actual *application* of the absorption line studies, as tools in understanding the nature of galaxies (and the intergalactic medium) as a function of time, are in a state of infancy in many respects. It is of course here that the promise of the whole field must be realized: the unique access to physical details of galaxies as a function of time must be exploited and then integrated into the global picture of galaxy formation and evolution.

The aim of this contribution is two–fold: first, in order to understand the overall implications *for galaxies* of the studies of metal line absorption systems, it is crucial to know exactly what types of galaxies represent potential absorbers, how the absorption properties might be related to the galaxy properties, and what parts of those galaxies give rise to various types of absorption signatures. Second, and very much related to the first (but perhaps of more immediate interest to the community interested in galaxy evolution *per se*), what can one infer about the evolution of normal field galaxies using the presence of absorption, rather than flux density in some observed bandpass, as the selection criterion? I will argue that, once the "selection function" of the absorbing galaxies is understood, absorption–selected galaxies can be used to understand the overall evolution of field galaxies to redshifts far beyond where an apparent magnitude limited sample breaks down, and can be used to examine the galaxy luminosity function in a potentially much less biased manner even in the redshift range of "overlap" between the methods.

From the point of view of those of us working in the field of QSO absorption lines, the most basic motivation for the kind of study described below is to provide a "context" in which all of the rich detail that will come out of the present and future absorption line studies can be understood and assimilated by those *outside* of the field. From the point of view of the galaxy evolution community, the methods represent an independent approach to examining the evolution of field galaxies with redshift, and the means to extend field galaxy "redshift surveys" (including, of course, the accompanying detailed physics accessible only through the absorption lines) well beyond $z = 1$. Because many of the results I will discuss have been summarized recently elsewhere (Steidel 1993, Steidel *et al* 1994, 1995, Steidel and Dickinson 1995), I will devote most of the discussion in this article to new results on the gaseous structure of absorbing galaxies and on the dependence of extended gaseous envelopes on galaxy properties.

## 2   Establishing the "Selection Function"

### 2.1   Observational Methods and Biases

As discussed in more detail elsewhere (see above references), we use as our absorption selection criterion the Mg II $\lambda\lambda 2796, 2803$ doublet (with rest equivalent width $W_0 > 0.3$Å) because the statistics are well-known for the entire redshift range $0.2 \leq z \leq 2.2$ (Steidel & Sargent 1992) and because it turns out to be a very effective tracer of H I gas that has $N(H\ I) \gtrsim 10^{17}$ cm$^{-2}$; for this reason we usually refer to the sample as being "gas cross-section selected" at an effective H I threshold that corresponds to gas with $\tau \gtrsim 1$ in the Lyman continuum (and so the sample could equally well be described as selected by the presence of a Lyman limit system [LLS]). It turns out that the redshift path density $dN/dz$ for such systems is consistent with no evolution in co-moving total cross-section (i.e., the product of the space density of the absorbers and the effective mean cross-section of individual absorbers remains constant) over the whole redshift range observed; this fact becomes very useful in deducing, e.g., luminosity functions and space densities of the galaxies as a function of redshift (see §2.5). The distinct advantages of an absorption-selected galaxy sample (from the point of view of understanding field galaxy evolution) are detailed in Steidel *et al* 1994.

Before summarizing the results of our completed survey for absorption-selected galaxies at $z \leq 1$, it is worth making some general comments on the technique used and possible biases inherent in the method. We have adopted a strategy similar in some respects to the one used in the seminal study of Mg II absorbing galaxies by Bergeron & Boissé (1991) – we obtain deep continuum images (for our survey, they are taken over a very wide wavelength baseline, from the visual to the K-band) of the fields of the QSOs and essentially begin our search for the absorber at the position of the QSO and work our way out in angular separation.

We wish to emphasize that, provided one obtains spectra of the candidate galaxies near the QSO sight line, the search strategy adopted is in no way biased against discovering non–absorbing galaxies; in other words, one is just as likely to identify a putative population of non-absorbing galaxies by searching in fields in which absorption is known to be present as where it is not – if they are there, they will be found. Nevertheless, in our study we have included with our sample of 51 absorber fields (chosen to have Mg II absorption with $W_0 \geq 0.3$ Å and $0.3 \leq z \leq 1.0$) about 25 QSO fields in which no Mg II absorption is present over the same range of redshift. These "control" fields have the obvious advantage that if no objects are seen close to the line of sight, no galaxy spectroscopy is necessary to resolve any confusion about the presence of an "interloper" (i.e., a galaxy close to the line of sight but producing no absorption). For the faint galaxy spectroscopy part of the program, we have made every attempt to obtain spectra of all detected objects within $\sim 10''$ of the QSO, corresponding to $\sim 50h^{-1}$ kpc at the typical redshift of the survey, and often we have obtained spectra of several objects per field at larger angular separations. This search radius was adopted largely from experience; if it had turned out to be necessary to go to larger radii to find a galaxy at the absorption line redshift, then we would have extended the search. What we have found is the nearest galaxy with a redshift matching that of the absorption system, and in several cases we have found evidence for galaxy groups (not surprising, since most present–day galaxies reside in groups). In any individual case, therefore, one cannot *prove* that the identified object is the absorber–it could be a galaxy at the same redshift but larger impact parameter. However, as will be described below, we see very significant correlations of the *identified* galaxy properties with the absorption line properties, and a remarkable absence of interloper galaxies at redshifts different from that of the absorption, so that it is clear beyond almost all reasonable doubt that the adopted strategy has succeeded in finding the actual absorber. Similarly, we cannot go infinitely faint in either the imaging or the spectroscopy, so that the possibility would always exist that the "true" absorber is a hopelessly faint object, perhaps situated directly on top of the QSO (see, e.g., York *et al.* 1986, Yanny & York 1992); however, this too can be rejected in a statistical sense because we have found a galaxy at the right redshift in every case *without* reaching extremely faint levels (although the identified galaxies do in fact represent a very wide range in absolute luminosity), and as will be detailed below, the galaxy luminosities, the distribution of projected impact parameters, and the observed absorption line characteristics form a remarkably coherent relationship.

## 2.2  The Nature of the Absorbing Galaxies, $z \leq 1$

As stated in §1, many of the results of the survey have been presented elsewhere, and for lack of space in the current contribution, we refer the interested reader to those references for more details. We concern ourselves here primarily with what types of galaxies are represented in a sample selected by metal line absorption. It has turned out that the K–band data has been extremely important to an interpretation of the results of the absorbing galaxy survey; initially the purpose of the near–IR data was to get an idea about the extent to which the observed rest–frame B band luminosities constructed from our optical data[2] were biased by current rates of star formation. It turns out, however (see §2.3), that it is the $K$ luminosity that is a much more robust predictor of a galaxy's potential to produce detectable absorption. The wide color baseline also allows one to classify galaxy "morphological" (really, spectroscopic) types, which are extremely useful in the absence of $HST$–resolution images for every field.

First, the widely cited conclusion of Bergeron & Boissé (1991) that only "bright" galaxies are absorbers turns out not to be completely accurate in view of a much larger sample; while the absorbers have a mean luminosity of about $0.7L_B^*$ and $0.5L_K^*$ (where these refer to *rest–frame* luminosities), the actual range of luminosity spans more than a factor of 70, with the faintest identified absorber having $L_K \approx 0.05L_K^*$. The spectra and the optical/IR colors of the absorbing galaxies suggest that the sample is drawn from the *full range of galaxy spectroscopic types*, from very blue objects with the colors or present–day late–type spirals and Im galaxies, to objects which have the colors and spectra of completely unevolved elliptical galaxies (see Figure 1). There is no evidence for any tendency for the absorbing galaxies to have particularly high star formation rates, on average. As can be seen from Figure 1, the "average" absorber maintains the rest–frame $B - K$ color of an Sb spiral for the full range of redshifts observed. This is consistent with a $\sim$ constant rate of star formation for the absorbing galaxy population as a whole since at least $z \sim 1$. We also see no evidence for luminosity evolution of the population (in either rest–frame B or rest–frame K) over the same redshift range (see Steidel *et al.* 1994, Steidel & Dickinson 1995). One of the most surprising conclusions (see also §2.3 below) is that galaxies of similar $K$ luminosities but with widely differing current star formation rates appear to have extended gaseous envelopes that are indistinguishable. The implications of this fact will be discussed in §4.

---

[2]  It is important for the purposes of obtaining accurate luminosity distributions to make the appropriate k–corrections to reach rest–frame absolute magnitudes; for example, simply using an observed $R$ magnitude with no bandpass correction to obtain M(R) can lead to relative errors of up to 1 magnitude for galaxies in the redshift range of interest.

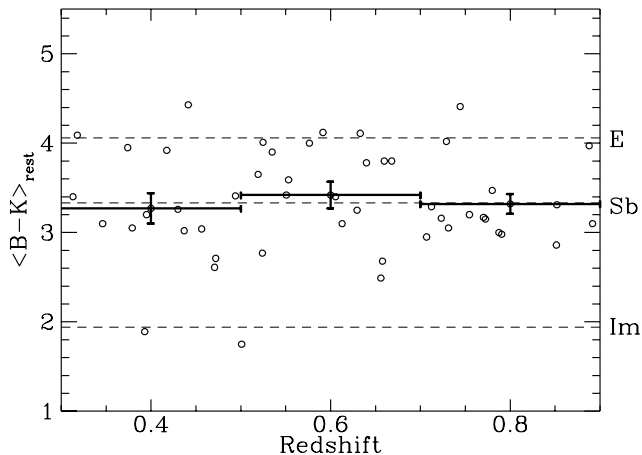

**Fig. 1.** Plot showing the distribution of rest–frame optical/IR colors of the absorbing galaxies versus redshift. Note that the galaxies appear to represent the full range of spectroscopic types, but have an average color equivalent to a present–day Sb galaxy across the whole redshift range.

### 2.3 The Geometry of the Extended Gas and the Absorber Selection Function

The advantage of a large sample of absorbing galaxies is that it begins to become feasible to extract significant relationships between the gaseous sizes of the galaxies, the galaxy luminosity, and the absorption line properties. Establishing these statistical trends, together with the determination of the nature of any non–absorbing population, is crucial to inserting the absorbing galaxy population into the context of general field galaxy evolution. For example, it has been somewhat of a "tradition" over the last ∼ 20 years to calculate the expected "sizes" of QSO absorbers of various classes based on knowledge of $dN/dz$ and assumptions about the galaxy luminosity function, luminosity/ size scaling relation, and gas–phase geometry. It is now possible to turn the whole problem around, and *infer* (based on direct observations) all of the relations that were always assumed, and directly derive the galaxy luminosity function. If we understand what selection function is in effect for gas cross-section selected galaxies, it is possible to produce a field galaxy luminosity function directly that is almost completely independent of those produced from the apparent magnitude selected surveys of recent years (see, e.g., Lilly 1993, Songaila *et al.* 1994, Colless 1995).

The usual assumption has been that the gaseous sizes of galaxies would obey the Holmberg (1975) relation, $R_{\rm gal} \propto L^{0.4}$, and the luminosities are drawn from a Schechter (1976) luminosity function. On the basis of a subset of the current sample, we made the claim that the absorbers followed a relation that is closer to $R_{\rm gal} \propto L^{0.2}$ (Steidel 1993); this exponent, often called $\beta$, is a very important number to determine because it controls the extent to which the gas cross-section selection criterion depends upon luminosity, and hence it determines the effective volume probed by an absorption line survey as a function of galaxy luminosity. We have undertaken a new determination of $\beta$, using the following method:

We include all galaxies with redshifts (whether they have produced absorption or not) in the survey fields. We then assume spherical "halos" of gas, obeying the relation $R = R^*(L/L^*)^\beta$, searching for the best values of $R^*$ and $\beta$, subject to the following criteria. First, we minimize the number of galaxies that fall on the "wrong" side of the $R-L$ curve (i.e., minimize the number of absorbing galaxies falling outside the curve and the non–absorbing galaxies falling inside the curve), and we also insist that the distribution of normalized impact parameters ($D_n = D_i/R_i(L)$, where $D_i$ is the observed impact parameter and $R_i(L)$ is the expected halo extent for a galaxy of the same luminosity as the observed galaxy) is consistent with the expected radial distribution, which should be $n(D_n)dD_n \propto D_n$ for the assumed geometry. This fitting process results in two slightly different relationships, $R(L_K) = 38h^{-1}(L_K/L_K^*)^{0.15}$ and $R(L_B) = 35h^{-1}(L_B/L_B^*)^{0.2}$, with the former having smaller scatter. The halo size–K luminosity relationship is plotted in Figure 2; the most important point to notice is how amazingly well the simple model actually describes the data (and how poorly a model with $\beta = 0.4$ fits the data!). Note that under the best–fit curve there are essentially *no non–absorbing galaxies*. This results in two very important conclusions: first, if a galaxy is brighter than $M_K = -22$ (or about $0.06 L_K^*$) and it falls within $R(L_K)$ of a QSO sight line, it will produce detectable Mg II absorption – this means that inside that radius, the covering fraction of the absorbing gas is unity, independent of galaxy spectroscopic/morphological type. Secondly, this same absence of non–absorbers below the curve means that the assumption of spherical halos *must* be close to the truth. For example, if the absorption were all produced by extended thin disks, one would expect, statistically, that 50% of the galaxies within $R(L_K)$ would produce no detectable absorption (see also §2.6 below).

On the other hand, despite the surprising "regularity" of the above results, there is evidence that the existence of extended gas capable of producing absorption depends on environment, at least in extreme cases. Bechtold & Ellingson (1992) have shown that galaxies located in the same clusters as QSOs generally do not produce detectable MgII absorption. However, cluster cores have a very small cross-section on the sky, so it is not surprising that, in a collection of 100 or so random lines of sight, we did not probe any such environments. Again, we already have evidence that many of the absorbers are members of groups, but there is clearly much work to be done in estab-lishing the detailed environments of the galaxies and any possible effect on the absorbing cross-sections or halo gas kinematics.

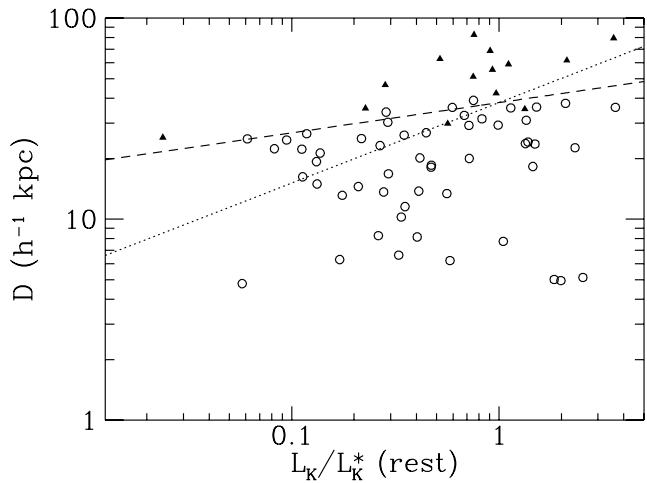

**Fig. 2.** Plot showing observed impact parameter $D$ versus galaxy near-IR luminosity $L_K$ for the absorbing (open circles) and non–absorbing (filled triangles) galaxies. The dashed curve is the best–fit relation with $\beta = 0.15$ and $R^* = 38h^{-1}$ kpc, as described in the text. The dotted curve has $\beta = 0.4$, which clearly fails badly in describing the data.

### 2.4 Dependence of Absorption Line Properties on Galactocentric Distance

Lanzetta and Bowen (1990) have noted a significant inverse correlation between the observed equivalent width of the Mg II absorption lines and the impact parameter between the QSO sight line and the centroid of the absorbing galaxy from the sample of Bergeron & Boissé (1991). Since it is fairly well-established from high–resolution observations that the Mg II equivalent width is proportional to the number of individual velocity components (e.g., Petit-Jean & Bergeron 1990), it is natural to try to obtain a radial "cloud" density distribution function from the observed correlation. Unfortunately, a larger data set reveals that the situation is probably much more complicated; the data for our $z \leq 1$ sample are plotted in Figure 3. Note that most, if not all, of the very high equivalent width systems turn out to be damped Lyman alpha absorbers, with impact parameters smaller than $15h^{-1}$ kpc.[3] Most of

---

[3] We have no direct information on the H I column density for most of the $z < 1$ absorbers; note also that many of the very strong systems are taken from the

the correlation is produced by these systems, where galactic rotation almost certainly enters into the observed kinematics (see Art Wolfe's contribution to these proceedings). Note also that *all* of the known damped systems have $D \leq 15h^{-1}$ kpc, but that several of them have very small Mg II equivalent widths (some of them would not make it into a Mg II selected sample with $W_0 > 0.3$Å ). Figure 3 emphasizes the need to explore the detailed relation of the absorption line kinematics to the actual galaxy morphology and sky orientation; this can be done to $z \sim 1$ with HST images in combination with high-resolution spectroscopy (see §2.6).

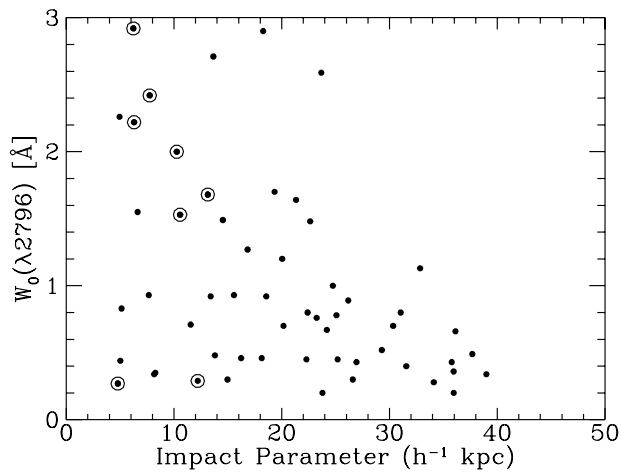

**Fig. 3.** Plot showing the Mg II equivalent width versus impact parameter. Points surrounded by circles are known damped Lyman $\alpha$ absorbers. The anti-correlation of line strength and impact parameter is significant at the $3.1\sigma$ level.

### 2.5   The Galaxy Luminosity Function

Armed with a very robust empirical determination of the scaling of galaxy gas cross-sections with luminosity, one can then turn the observed distribution of galaxy luminosities into a true luminosity distribution (provided that the absorption systems themselves have been drawn from a statistically homogeneous sample) simply by weighting the histogram by $1/R^2(L)$, which for the $K$ luminosity would be $(L_K/L_K^*)^{-0.3}$. This cross-section weighting is equivalent to dividing by the effective "volume" probed by an absorption line survey

---

literature and are *not* part of our unbiased sample (although we have re-observed all of them as part of our survey). In a sample the size of ours ($\sim 60$ absorbers) a homogeneous sample would contain only 1 or 2 systems with $W_0 > 2$ Å.

as a function of luminosity, directly analogous to the selection function in an apparent magnitude selected redshift survey. However, note that the volume probed in an absorption-selected survey is much more weakly dependent on luminosity ($L^{0.3}$ rather than $L^{1.5}$!) and therefore one is actually sampling the full range in galaxy luminosities at a given redshift much more uniformly in an absorption–selected sample. The absorbing galaxy K–band luminosity function is is shown in Figure 4; note that it is quite well-represented by a Schechter luminosity function with faint end slope $\alpha \approx -1$, to $M_K \approx -22$. The absolute normalization of the luminosity function is obtained from the same relationship that has traditionally been used to calculate $R^*$ from an assumed luminosity function and scaling relation,

$$\frac{dN}{dz} = \frac{c\sigma n}{H_0}(1+z)(1+2q_0 z)^{-0.5}$$

where

$$\sigma n = \pi \int_{L_{min}}^{\infty} \Phi(L) R^2(L) dL.$$

Here $dN/dz$, $R(L)$, and $L_{min}$ are empirically determined, and $\Phi^*$ is calculated by assuming that a Schechter function which fits the observed luminosity distribution applies.

We note that the shape of the rest–frame B luminosity function is quite different (see Steidel *et al.* 1994, Steidel & Dickinson 1995), resembling a Gaussian rather than a Schechter function. We have attributed this difference to the fact that the population of intrinsically small (faint $M_K$), star forming "faint blue galaxies" (which contribute significantly to the field galaxy luminosity function even at relatively bright values of $M_B$) apparently do not possess significant gaseous envelopes. This is further evidence that the existence of the gaseous "halos" is dependent on galaxy *mass*, independent of star formation rate. In fact, we can see from the absorber luminosity function that the luminosity/cross-section scaling relation derived above *must* break down for luminosities smaller than $L \approx 0.05 L_K$ – otherwise we would have found either many more very faint absorbers (the relatively weak dependence on luminosity and the knowledge that the luminosity function is rising at faint magnitudes requires this) or more "blank" fields. More work that explores the "transition" region of the luminosity function (at faint intrinsic luminosities), from absorber to non–absorber, would be very important in establishing the nature/origin of the absorbing gas.

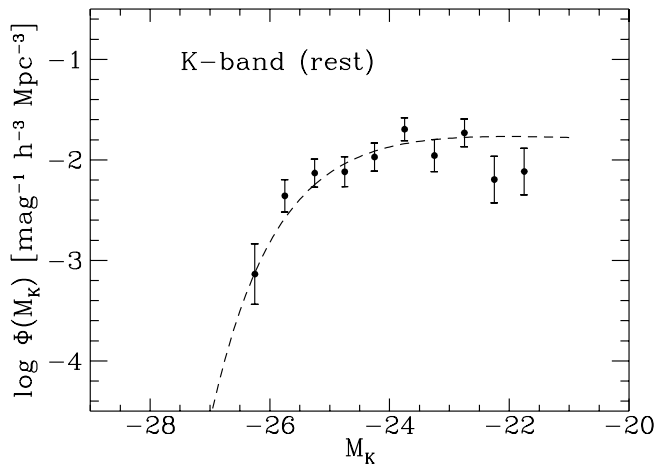

**Fig. 4.** Plot of the absorbing galaxy K–band luminosity function for the sample with $z \leq 1$ ($\langle z \rangle = 0.65$). A recent determination of the local field galaxy K-band luminosity function (Mobasher et al. 1993) is also shown (with arbitrary normalization).

### 2.6  Some Preliminary HST Imaging Results

We (Steidel, Dickinson, Meyer, & Sembach) have obtained a very deep (24,000 s) image with WFPC-2 on board HST of the field of the QSO 3C 336, which has 5 absorbing galaxies along a single line of sight, allowing us to directly observe the galaxy morphology and orientations with respect to the QSO sight line. First of all, we are very encouraged that the actual observed morphological types match very well the types we had assigned the galaxies on the basis of the spectra and optical/IR colors, in this field ranging from Sd/Im to S0. The $z = 0.892$ absorber is clearly a nearly edge-on mid-type spiral, but it also produces a Lyman $\alpha$ absorption line with $N(H\ I) = 5 \times 10^{19}$ cm$^{-2}$ on the basis of our FOS spectrum. The galaxy has a projected separation from the line of sight of $15h^{-1}$ kpc, but after correction for inclination, if one insists that the absorption arises in the disk, the minimum disk extent would have to be $\sim 70h^{-1}$ kpc, in a galaxy with a luminosity of only $0.3L^*$. Despite intensive spectroscopy, the identification of the $z_{abs} = 0.656$ system, which produces a damped Lyman alpha line with $N(H\ I) = 2 \times 10^{20}$ cm$^{-2}$, remains ambiguous; the galaxy responsible must be considerably fainter than $\sim 0.1L^*$. This may serve to caution us against assuming that all high column density gas must be confined to disks of luminous galaxies.[4]

The 5 galaxies producing metal lines along this sight line have impact parameters ranging from $8 - 47h^{-1}$ kpc, and on the basis of our FOS spectra there is a clear trend for the equivalent width ratio W(C IV)/W(Mg II) to increase with galactocentric distance. More data of this kind will clearly be interesting in investigating the actual ionization structure of the halos [see also Bergeron (1995), Lanzetta (1995)].

## 3  Absorbing Galaxies at Higher $z$

### 3.1  $z > 1$

One of the advantages of an absorption–selected sample is that the galaxies can be followed, using identical selection criteria, well beyond $z = 1$, where faint apparent magnitudes for objects of normal luminosity and lack of strong spectroscopic features cause conventional field galaxy surveys to break down. We have recently begun a survey designed to follow the absorption-selected galaxies to $z \sim 1.6$ (see Steidel & Dickinson 1995 for details) in order to establish the field galaxy luminosity function (using techniques described above) at $\langle z \rangle = 1.3$. Insisting on spectroscopic completeness would of course encounter the same difficulties as the field galaxy redshift surveys; however, for the absorption selected sample we make use of very strong predictions based on the $z < 1$ sample (on color, angular separation, apparent magnitude, etc.), together with *a priori* redshift information so that evolution of the galaxy population producing absorption can be evaluated using deep imaging (one optical passband and one near-IR (K) passband) alone. This allows us to establish the galaxy luminosity function over a significant range of intrinsic luminosity in redshift regimes that are unexplored by the field galaxy redshift surveys. Our preliminary results suggest that no spectacular evolution is occurring even to $z \sim 1.6$ in the field galaxy population, and in fact we have been able to establish that the space density of the normal Hubble sequence galaxies (that appear to dominate the absorption cross-section) remains constant to within $\sim 30\%$ (Steidel & Dickinson 1995), and that there has been at most modest passive evolution of the stellar populations (here again the K–band data are crucial for epoch-to-epoch comparisons) over that entire redshift range. Taken together with the results of the field galaxy redshift surveys, a scenario in which relatively luminous galaxies have been in place since very high redshifts, but where the population of very blue galaxies is undergoing substantial evolution even at relatively modest redshifts, seems to be indicated. A great deal of work is called for in understanding the possibly intimate connection between the "halo" gas supply of the luminous galaxies (which we see in absorption) and the rapidly evolving "faint blue galaxy" population (see §4).

---

[4] We also note that the *known* damped Lyman $\alpha$ systems in our $z < 1$ sample show a marginally significant tendency to be fainter and bluer than the Mg II selected sample as a whole. The mean luminosity is only $\sim 0.3 L^*$.

### 3.2   $z > 3$

Absorption line selection has clear advantages for "flagging" the sites of known very high redshift objects; since LLSs are now known far beyond $z = 4$, in principle the same selection criteria can be used at these extreme redshifts as at lower redshift to target searches for the absorbing galaxies. At $z > 3$, one can take advantage of a *guaranteed* spectroscopic feature in the spectrum of any object forming stars–a discontinuity at the rest frame Lyman limit of the galaxy, redshifted into the optical portion of the spectrum. One need not make any assumptions about the fraction of ionizing photons escaping a young galaxy, or even the details of the intrinsic spectral energy distribution of the galaxy, as a pronounced discontinuity will be present from consideration of the opacity of the IGM (Lyman $\alpha$ forest blanketing plus optically thick intervening clouds) alone (see Madau 1995). We were able to take advantage of the expected characteristic colors of objects in the redshift range $3 \leq z \leq 3.5$ in a custom photometric system, $U_n G \mathcal{R}$, to identify the object which may be producing the damped Lyman $\alpha$ absorption system at $z = 3.39$ toward Q0000−263 (Steidel & Hamilton 1992), and possibly to identify a group or cluster of objects associated with it (Steidel & Hamilton 1993; Giavalisco *et al.* 1994). Max Pettini and I have been continuing this program in an effort to establish the detectability of other LLS-selected high redshift galaxies in the same redshift range (Steidel & Pettini 1995); in total, we have identified 2 out of 5 of the systems observed (including Q0000−263), and found many $z > 3$ galaxy candidates that may be associated with the absorbers or with the background QSOs. Both putative absorbing galaxies have luminosities only slightly in excess of $L^*$. The relatively low detection rate compared to the lower redshift surveys may be ascribed to the fact that our current detection limits only allows us to detect galaxies of luminosity $\gtrsim 2L^*$. At present the population of absorbers at $z > 3$ is largely unconstrained (and a luminosity distribution consistent with the lower $z$ absorbers cannot be ruled out), but observations going 1−2 magnitudes deeper (quite possible with 8−10 m telescopes) will be very telling. Also of interest, of course, is the galaxy environment and morphology at such early times. These can also be addressed currently using HST and large ground-based telescopes.

## 4  What Does It All Mean?

A general result, about which we can all rest assured at this point, is that the QSO metal line systems (at least, those that are optically thick in the Lyman continuum; see Ken Lanzetta's contribution for extensions to lower H I column density regimes) are produced by normal galaxies and therefore in studying the absorption line systems one is justified in claiming that we are learning about the history of such objects. However, the absorbers exhibit (at least to $z \sim 1.5$ or so) a very large range in star formation rate, color, luminosity, and from the limited data that exist, morphological type. This means that when we talk about evolution in metallicity, dust content, kinematics, etc., we are talking about a very wide range of galaxy types and we must keep this in mind – it may well be that the wide range in metallicity observed at $z \gtrsim 2$ (Pettini 1995) reflects very different evolutionary histories for various sub-populations of galaxies.

On the other hand, the one property of the of the galaxies that seems to be most important in determining the size and geometry of the absorbing gas is the galaxy *mass*, and the geometry and sizes inferred are remarkably regular in nature. The picture of spherical halos with a cutoff near $40 h^{-1}$ kpc is no doubt a tremendous over-simplification of the true situation; however, again, the data indicate that something is regulating the extended gas distribution to a very large extent. That the star formation rate does not appear to be important to the nature of the halos (I believe) argues very strongly against a model in which the extended gas is produced by outflow from the central regions (at least one which is powered by star formation, as in models of giant "galactic fountains"). Rather, the data seem to point to *inflow* as the dominant source of gas, and the relatively short timescales for the existence of gas in the extended halos (because of cloud–cloud collisions and dissipation) means that the inflow must occur continuously over at least a Hubble time (the time period over which we observe the absorption). What is a self-consistent physical picture that can explain the regularity of the gas at an effective H I column density threshold of $\sim 10^{17}$ cm$^{-2}$, and make it weakly dependent on mass alone? I suspect that the link is in the pressure (and pressure gradient) of the medium confining the $\sim 10^4$ K clouds, in which case there would be a characteristic radius at which the pressure is sufficient to make the density of the infalling material just optically thick in the Lyman continuum in the presence of the prevailing UV radiation field (whereas the gas distribution will extend, at lower column densities, to much larger galactocentric radii [see Lanzetta 1995]). This pressure, if it is due to some hot medium, is expected to be dependent only on the overall size of the potential well, hence the mass dependence and roughly spherical geometry. Whether or not this picture is correct, there is clearly a great deal of theoretical work to be done in understanding the physics of the absorbing regions.

One final comment: it is not just from the new generation of 8-10m telescopes and high resolution spectrographs that the major advances will come in our field. It is clear that the picture one can assemble from *combining* the information obtained from the QSO spectroscopy and from HST and ground–based imaging and spectroscopy of the same galaxies will allow us to overcome the limitations imposed by photon starvation and actually understand how *normal* galaxies have evolved, from the epoch of their formation to the present.

I would like to thank my collaborators Mark Dickinson, Dave Meyer, Eric Persson, Max Pettini, and Ken Sembach for allowing me to quote the results of joint work prior to publication.

# References

Bechtold, J., Ellingson, E. 1992, ApJ, 396, 20
Bergeron, J. 1995, this volume
Bergeron, J., Boissé, P. 1991, A&A, 243, 344
Colless, M. M. 1995, in Wide Field Spectroscopy and the Distant Universe, proc. of the 35th Herstmonceaux Conference, eds. S. Maddox and A. Aragon-Salamanca, in press.
Giavalisco, M., Steidel, C. C., Szalay, A. 1994, ApJL, 425, L5
Holmberg, E. 1975, in Stars and Stellar Systems, 9, Galaxies and the Universe, eds. A. Sandage, M. Sandage, and J. Kristian, (Chicago: University of Chicago Press), p 123.
Lanzetta, K. M. 1995, this volume
Lanzetta, K. M., Bowen, D. V. 1990, ApJ, 357, 321
Lilly, S. J. 1993, ApJ, 411, 501
Madau, P. 1995, ApJ, in press
Mobasher, B., Sharples, R. M., Ellis, R. S. 1993, MNRAS, 263, 560
Petit-Jean, P., Bergeron, J. 1990, A&A, 231, 309
Pettini, M. 1995, this volume
Schechter, P. 1976, ApJ, 293, 97
Songaila, A., Cowie, L. L., Hu, E. M., Gardner, J. P. 1994, ApJS, 94, 461
Steidel, C. C. 1993, in The Environment and Evolution of Galaxies, proc. of the 3rd Teton Astronomy Conference, eds. J. M. Shull and H. A. Thronson, (Dordrecht: Kluwer), p. 263
Steidel, C. C., Dickinson, M. 1995, in Wide Field Spectroscopy and the Distant Universe, proc. of the 35th Herstmonceaux Conference, eds. S. Maddox and A. Aragon-Salamanca, in press.
Steidel, C. C., Pettini, M. 1995, in preparation
Steidel, C. C., Dickinson, M., Persson, S. E. 1995, in preparation
Steidel, C. C., Dickinson, M., Persson, S. E. 1994, ApJL, 437, L75
Steidel, C. C., Hamilton, D. 1992, AJ, 104, 941
Steidel, C. C., Hamilton, D. 1993, AJ, 105, 2017
Steidel, C. C., Sargent, W. L. W. 1992, ApJS, 80, 1
Wolfe, A. M. 1995, this volume
Yanny, B., York, D. G. 1992, ApJ, 391, 569
York, D. G., Dopita, M., Green, R., Bechtold, J. 1986, ApJ, 311, 610